\documentclass [aps,prl,floats,showpacs,amsmath,twocolumn,superscriptaddress,10pt]{revtex4-1}

\usepackage{mathrsfs}
\usepackage{natbib}

\usepackage{graphicx}	
\usepackage[english]{babel}
\usepackage[T1]{fontenc} 
\usepackage[latin1]{inputenc}  
\usepackage[babel=true]{csquotes}
\usepackage{color}
\usepackage{bm}
\usepackage{import}
\usepackage{dcolumn}

\renewcommand{\vec}[1]{\mathbf{#1}}
\newcommand{\orient}{\theta}
\newcommand{\epsc}{\ensuremath{\varepsilon_c}}

\newcommand{\cd}{C_D}
\newcommand{\sizetick}{\small}

\newcommand{\spanw}{a}
\newcommand{\stream}{h}

\begin{document}
\title{Drag Reduction, from Bending to Pruning}
\author{Diego Lopez}
\email{diego.lopez@ladhyx.polytechnique.fr}
\altaffiliation{\\Current address: Dept. Mechanical Aerospace Engineering, University of California, San Diego, La Jolla, CA, USA}
\affiliation{LadHyX, D\'epartement de M\'ecanique, \'Ecole Polytechnique, 91128 Palaiseau, France}
\author{Christophe Eloy} 
\affiliation{Aix-Marseille University, IRPHE UMR 7342, CNRS, 13013 Marseille, France}
\affiliation{Department of Mechanical and Aerospace Engineering, University of California, San Diego, 9500 Gilman Drive, La Jolla, California 92093-0411, USA}
\author{S\'ebastien Michelin}
\affiliation{LadHyX, D\'epartement de M\'ecanique, \'Ecole Polytechnique, 91128 Palaiseau, France}
\author{Emmanuel de~Langre}
\affiliation{LadHyX, D\'epartement de M\'ecanique, \'Ecole Polytechnique, 91128 Palaiseau, France}
\date{\today}

\begin{abstract}
Most plants and benthic organisms have evolved efficient reconfiguration mechanisms to resist flow-induced loads. These mechanisms can be divided into bending, in which plants reduce their sail area through elastic deformation, and pruning, in which the loads are decreased through partial breakage of the structure.
In this Letter, we show by using idealized models that these two mechanisms or, in fact, any combination of the two, are  equally efficient to reduce the drag experienced by terrestrial and aquatic vegetation. 
\end{abstract}
\maketitle


A major mechanical constraint on terrestrial plants and benthic organisms results from  external fluid flows. To resist large flow-induced loads, plants have evolved two types of adaptability. The first one is a long time-scale adaptability, where the flow induces shape modifications of the plants \cite{moulia_2006,telewski_2006}. The second one may occur on long or short time-scales and involves time-reversible geometrical changes. This latter mechanism is known under the general term of reconfiguration \cite{vogel_1984,delangre_2008}, which can itself be  divided into two distinct mechanisms: bending and pruning.

In bending, also known as elastic reconfiguration, the body deforms significantly under the flow forces, thereby reducing its drag compared to that of a non-deformable body \cite{vogel_1989}. Over the past decade, several experimental, theoretical and numerical studies have provided a good understanding of this mechanism \cite{koehl_1984,kane_2008,alben_2002,gosselin_2010,delangre_2012a}.
Alternatively, in pruning, the sail area is reduced by breaking parts of the plant structure, as may be  observed in trees \cite{lopez_2011,eloy_2011}. 
Like bending, pruning leads to significant drag reduction. However, to ensure the survival of the organism and the long-time reversibility of this type of reconfiguration, breakage should be localized away from the base of the structure for the organism to be able to regrow. 

\begin{figure}[tb]
\newcommand{\testsize}{\footnotesize}
\newcommand{\xa}{\sizetick $10^{0}$}
\newcommand{\xb}{\sizetick$10^{1}$}
\newcommand{\xc}{\sizetick$10^{2}$}
\newcommand{\xd}{\sizetick$10^{3}$}
\newcommand{\xe}{\sizetick$10^{4}$}
\newcommand{\xg}{\sizetick$10^{5}$}
\newcommand{\xh}{\sizetick$10^{6}$}
\newcommand{\ya}{\sizetick$10^{-1}$}
\newcommand{\yb}{\sizetick$10^{0}$}
\newcommand{\yc}{\sizetick$10^{1}$}
\newcommand{\yd}{\sizetick$10^{2}$}
\newcommand{\ye}{\sizetick$10^{3}$}
\newcommand{\yg}{\sizetick$10^{4}$}
\newcommand{\yh}{\sizetick$10^{5}$}
	\centering
		\def\svgwidth{0.9\columnwidth}
		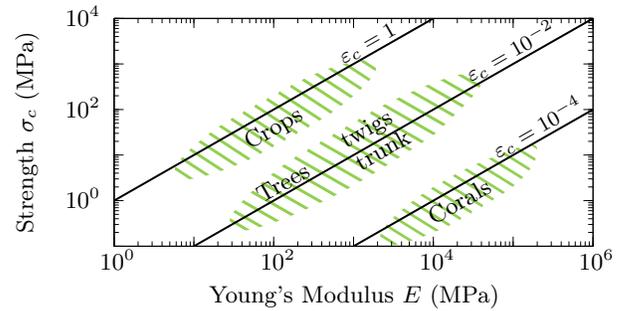
	\caption{Typical orders of magnitude of Young's modulus and strength in plants and benthic organisms \cite{crook_1994,baker_1995, beismann_2000, speck_2011, rosner_2007,*bjurhager_2008,*butler_2012,*gibson_2012,*note_corals}.}
	\label{fig:1}
\end{figure}

The deformation that an organism can withstand before eventually breaking can be assessed by comparing two material properties: the Young's modulus or modulus of elasticity, $E$, and the yield stress or strength $\sigma_c$. The former is the ratio between stress $\sigma$ and strain $\epsilon$ in the material, while the latter characterizes the stress at breakage. The dimensionless number formed out of these two quantities,  $\epsc=E/\sigma _c$, is simply the strain at breakage, or critical strain. 
Physically, a low value of $\epsc$ means that breakage will occur at small deformations.
As seen in Fig.~\ref{fig:1}, the critical strain, $\epsc$, can differ by several orders of magnitude in organic materials. Yet, terrestrial and benthic organisms are generally submitted to similar flow-induced loads. One may thus wonder how different values of $\epsc$ may affect their mechanical response and their ability to survive in intense flows. 

In this Letter, we explore how the critical strain may select different plant reconfiguration strategies (i.e. bending, pruning, or a combination of both) and show that all these strategies are equally efficient at reducing the flow-induced drag on idealized systems.

\begin{figure}[tb]
\footnotesize
	\centering
		\def\svgwidth{0.95\columnwidth}
		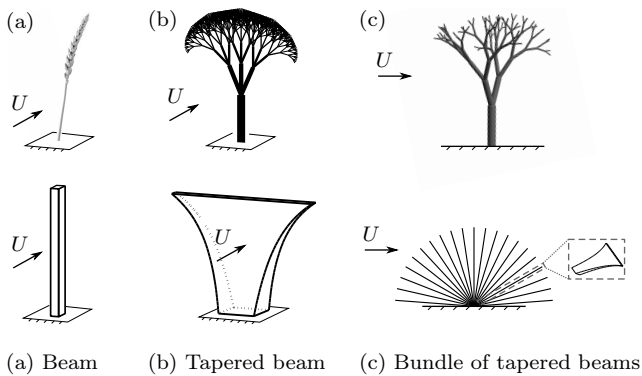
	\caption{Typical plant geometries and corresponding models: (a) beam (no ramifications) \cite{baker_1995,luhar_2011}, (b) Tapered beam (ramifications, no branch orientation) \cite{macmahon_1975, eloy_2011}, (c) bundle of tapered beams (ramifications and branch orientation) \cite{gosselin_2011}.}
	\label{fig:2}
\end{figure}

Plants, and more generally slender organisms under flow, are modeled here as cantilever tapered beams,  whose width, $\spanw$, and thickness, $\stream$, vary as 
\begin{equation}
	\spanw(s) = \spanw_0 \left(s/s_1\right)^{-\beta},\quad \stream(s)= \spanw_0 \left(s/s_1\right)^{\beta},
\end{equation}
where $s$ is the coordinate along the beam axis measured from top ($s=s_0$) to bottom ($s=s_1$), and $\beta$ is the slenderness exponent \cite{macmahon_1975,eloy_2011}.
When $\beta=0$, the beam has constant width and thickness along its height and represents a single-beam plant, such as a cereal crop (Fig.~\ref{fig:2}a). 
When $\beta >0$, the tapered beam represents a ramified system (Fig.~\ref{fig:2}b). In that case, the thickness, $\stream$, corresponds to the local branch diameter, and the width, $\spanw$, to the diameter multiplied by the number of branches at that height \cite{macmahon_1975}. The width thus corresponds to the total area facing the flow, or the sail area, thereby neglecting any shading between branches and the effect of branch orientation.
To investigate the impact of branch orientation on our results, we finally consider a ``bundle'' made of several tapered beams of different orientations (Fig.~\ref{fig:2}c).

\begin{figure}[tb]
\footnotesize
	\centering
		\def\svgwidth{0.8\columnwidth}
		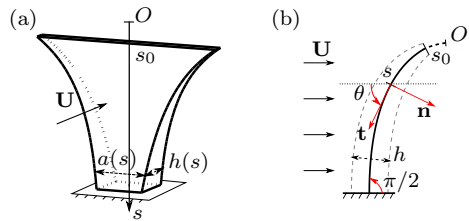
	\caption{Tapered beam and notations.}
	\label{fig:3}
\end{figure}

These tapered beams are assumed to lie in a uniform cross-flow of velocity $U$ (Fig.~\ref{fig:3}) such that flow-induced loads (force per unit length), $\vec{F}$, resulting from pressure drag read
\begin{equation}
	\vec{F} (s)= F\,\vec{n}, \quad \text{with } F=\tfrac{1}{2}\rho \cd \spanw \left[U \sin\orient\right]^2,
	\label{eq:part2_fflu}
\end{equation}
where $\rho$ is the fluid density, $\orient(s)$ is the local beam orientation, and $\cd$ is the drag coefficient, taken to be 1 without loss of generality \cite{delangre_2008}. 

Considering the high slenderness of this geometry, beam theory can be used to derive the deflection and stress state along the beam axis (Fig.~\ref{fig:3}b). Noting $F_t$ and $F_n$ the tangential and normal components of the internal force, respectively, $I$ the beam's second moment of inertia, and $E$ its Young's modulus, the Euler-Bernoulli beam equation yields
\begin{align}
	\label{eq:p2_vn}
	F_n' + \orient'F_t + F 	& =  0,\\
	\label{eq:p2_vt}
  	F_t' - \orient'F_n 		& =  0, \\
	\label{eq:p2_th}
	(EI \orient')' + F_n 		& =  0,
\end{align}
where primes stand for differentiation with respect to $s$ \cite{salencon_2001}. Equations \eqref{eq:p2_vn}--\eqref{eq:p2_th} describe the evolution of internal forces and torques along $s$, and are solved numerically using a shooting method with a clamped boundary condition at the base, i.e. $\orient(s_1)=\pi/2$, and free boundary conditions at the top, i.e. $F_t(s_0)=F_n(s_0)=0$. 

In this system, the bending stresses are maximum at the surface of the beam, being  compressive (resp. extensive) on the inner side (resp. outer side). The maximum stress is given by $\sigma_{\textrm{max}} = E \orient' \stream/2$ \cite{niklas_1992}, and we assume that breakage occurs when and where $\sigma_{\textrm{max}} = \sigma_c$.  Following a breaking event, the broken part is removed, and the problem is then solved on the remaining geometry characterized by a different $s_0$.

Before studying the reconfiguration of a single tapered beam for any value of $\epsc$, it is worth investigating two asymptotic limits: brittle  materials characterized by very small critical strains, $\epsc \ll 1$ (these materials break before any significant bending), and highly deformable materials for which $\epsc \gg 1$ (essentially, these materials never break). 

In the classical limit of an elastic material without breaking, a common behavior exists for all structures made of plates in bending, beams in bending, or any combination thereof \cite{delangre_2012a} : for a  large deformation of the structure drag will increase as $U^{4/3}$, showing a significant drag reduction in comparison with the expected $U^2$ scaling for a rigid structure. This $4/3$  exponent can be derived form a simple scaling argument, considering that the original length scale, the size of the structure, does not contribute to the drag, because of the significant change of geometry through bending \cite{gosselin_2010}. This  drag reduction is consistent with experimental data on many plant species, aquatic or aerial \cite{delangre_2012a}. The tapering of the beam, considered here through the parameter $\beta$, is not expected to play any role in the drag exponent, the length  scale of variation of $a$ and $h$ being the size of the beam.  We therefore expect that   in the limit of large critical strain, $\epsc \gg 1$,  drag will vary as $U^{4/3}$ , when  the beam is  highly  bent. 

Conversely,  we may consider the limit of a beam made of a brittle material, $\epsc \ll 1$, solved in  Ref.~\cite{lopez_2011}. Two cases are found depending on the value of the slenderness exponent. If $\beta < 1$,  the flow-induced stress is always maximum at the base, and breaking will systematically occur there. Drag increases as $U^2$ before breaking, when it drops to zero.  When   $\beta > 1$, breaking occurs at an intermediate level, resulting in a     succession of breaking events as  the flow  velocity is increased. The corresponding drag shows    sudden reductions, at the breaking events, separated by increases in drag that are  quadratic with velocity.  This step by step process was described as flow-induced  pruning, and is relevant for  complex tree geometries. It  is very efficient in limiting drag, and is somehow reversible as broken parts may regrow. The condition on $\beta$ was shown to be compatible with data on trees  \cite{lopez_2011}. 

\

In the general case, when both bending and pruning are involved, it should be noted that, since the width of the tapered beam diverges as $s$ tends to 0,  flow-induced loads are  predominantly exerted on the upper part of the structure. As a consequence, the cut-off length $s_0$ plays a role in scaling the flow-induced loads. In branching structures such as trees, the ratio $s_0/s_1$ is given by the relative length of the last branches, and can safely be estimated to lie in the interval $10^{-3}<s_0/s_1<10^{-1}$ for most organic branched structures \cite{macmahon_1976}. In the following, we consider a typical value of this ratio $s_0/s_1=10^{-2}$.

We now investigate the reconfiguration through bending and pruning of a single tapered beam when the flow velocity is gradually increased. We focus here on the case $\beta >1$ for which intermediate breaking events (i.e. pruning) are possible. A typical value  is $\beta=3/2$  as  proposed by McMahon and Kronauer \cite{macmahon_1976} based on the principle of elastic similarity in trees. The case $\beta<1$, for which breaking systematically occurs at the base of the structure, will be discussed further at the end of this Letter.

\begin{figure}[tb]
\newcommand{\xa}{\sizetick $10^{-2}$}
\newcommand{\xb}{\sizetick$10^{-1}$}
\newcommand{\xc}{\sizetick$10^{0}$}
\newcommand{\xd}{\sizetick$10^{1}$}
\newcommand{\xe}{\sizetick$10^{2}$}
\newcommand{\xf}{\sizetick$10^{1}$}
\newcommand{\ya}{\sizetick$10^{-2}$}
\newcommand{\yb}{\sizetick$10^0$}
\newcommand{\yc}{\sizetick$10^2$}
	\centering
		\def\svgwidth{0.75\columnwidth}
		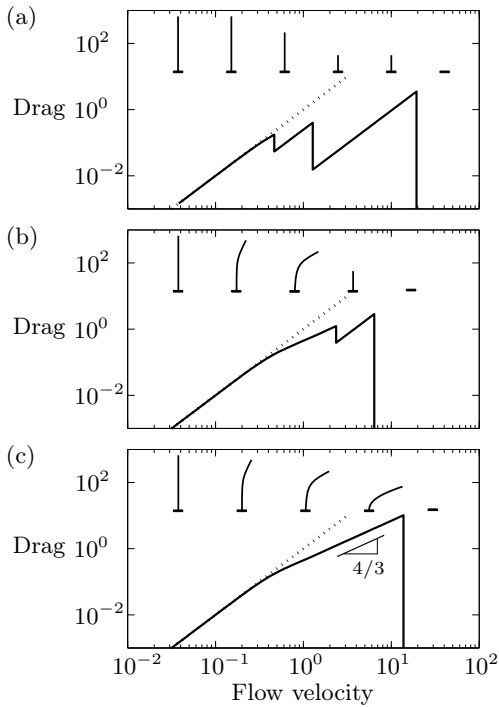
	\caption{Evolution of the normalized drag as a function of the flow velocity, for a tapered beam ($\beta=3/2$): (a) pruning reconfiguration, $\epsc = 10^{-3}$; (b) bending and pruning, $\epsc = 10^{-2}$; (c) bending reconfiguration, $\epsc = 10^{-1}$. The drag evolution in the absence of reconfiguration is shown in dotted line, and the beam topologies are sketched for different flow velocities.}
	\label{fig:4}
\end{figure}

Figure~\ref{fig:4} shows the computed evolution of drag with the flow velocity, for three values of the critical strain $\epsc=10^{-3}$, $10^{-2}$, and $10^{-1}$. In this figure, drag and velocity are normalized such that the drag is $1$ when the flow velocity is $1$ in the absence of reconfiguration.
For each value of the critical strain, $\epsc$, we see a significant drag reduction compared to that of a rigid body but with different mechanisms.
At low critical strain ($\epsc=10^{-3}$), little bending occurs, and reconfiguration is essentially due to pruning (Fig.~\ref{fig:4}a). This case is representative of fragile materials for which $\epsc \ll 1$. Pruning events are evidenced in Fig.~\ref{fig:4}a by sudden drops of the drag. 
Conversely, at large critical strain, here $\epsc=10^{-1}$, reconfiguration is essentially driven by bending, until pruning  eventually occurs through a single breaking event, at the base of the structure (Fig.~\ref{fig:4}c). This regime is reminiscent of highly deformable materials , $\epsc \gg 1$,  and  the finite value of $\epsc$ selects the flow velocity at which the structure breaks. 
Before breaking occurs, the asymptotic  bending regime of drag reduction described above is found, where drag   varies as $U^{4/3}$. 

In the intermediate case, $\epsc=10^{-2}$, reconfiguration and drag reduction are achieved through both bending and pruning  (Fig.~\ref{fig:4}b). Bending occurs prior to pruning, and therefore acts as a transition from no reconfiguration to pruning.

From the reconfiguration of a single tapered beam, we have seen that both bending and pruning  yield important drag reductions. Now, to assess the importance of branch orientations, we  model the reconfiguration of a bundle of such beams (Fig.~\ref{fig:2}c).
This geometry, inspired from the poroelastic system of \cite{gosselin_2011}, can be seen as a model for bushes or tree crowns, as each beam can itself be interpreted as a ramified branch.
Note that we consider here the worst case scenario, without any shading effect inside the bundle, thus assuming that each beam experience the same flow velocity $U$. 

\begin{figure}[t]
\newcommand{\xa}{\sizetick \ensuremath{10^{-1}}}
\newcommand{\xb}{\sizetick \ensuremath{10^{0}}}
\newcommand{\xc}{\sizetick \ensuremath{10^{1}}}
\newcommand{\xd}{\sizetick \ensuremath{10^{2}}}
\newcommand{\ya}{\sizetick \ensuremath{10^{-2}}}
\newcommand{\yb}{\sizetick \ensuremath{10^{0}}}
\newcommand{\yc}{\sizetick \ensuremath{10^{2}}}
\newcommand{\yd}{\sizetick \ensuremath{10^{1}}}
\newcommand{\ye}{\sizetick \ensuremath{10^{2}}}
\newcommand{\yg}{\sizetick \ensuremath{10^{-1}}}
\newcommand{\yh}{\sizetick \ensuremath{10^{1}}}
	\centering
		\def\svgwidth{0.75\columnwidth}
		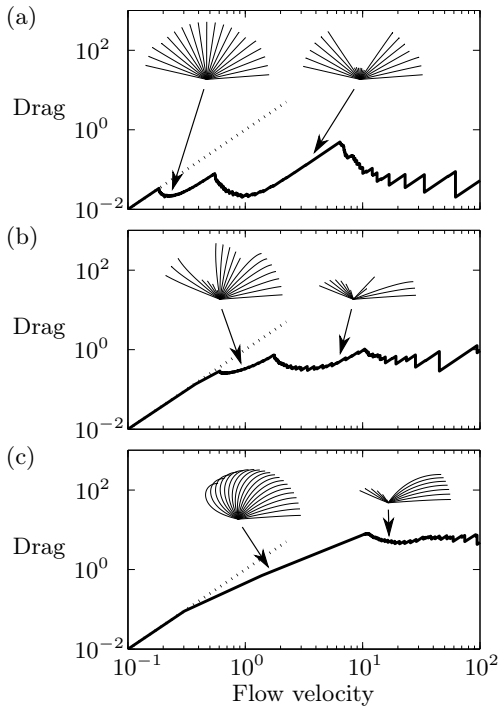
	\caption{Evolution of the normalized drag as a function of the flow velocity, for a radial bundle of tapered beams ($\beta=3/2$): (a) $\epsc = 10^{-3}$, (b) $\epsc = 10^{-2}$ and (c) $\epsc = 10^{-1}$. The drag evolution in the absence of reconfiguration is shown in dotted line, and the bundle topologies are sketched in different reconfigured states.}
	\label{fig:5}
\end{figure}

Figure~\ref{fig:5} shows the evolution of the drag on the bundle geometry when flow velocity is increased. 
Depending on the value of the critical strain, the topology of the reconfiguration is quite different. 
For low critical strain ($\epsc=10^{-3}$), breakage propagates from the center of the bundle to its periphery. 
Conversely, for high critical strain ($\epsc=10^{-1}$), breakage propagates downstream. 
At intermediate values of the critical strain ($\epsc=10^{-2}$), the behavior is a combination of these two propagative scenarios, leading to non-trivial reconfiguration patterns.
Yet, the different curves exhibit similar trends, and the resulting drag reduction is comparable for all values of the critical strain:  the total drag on the body is approximatively constant with the flow velocity, regardless of the type of reconfiguration involved. This confirms the results found for a single beam, suggesting that any value of the critical strain will lead to similar survival abilities under extreme flows.  
Furthermore, independently of the mechanism involved in the reconfiguration process, the combination of identical elements with different orientations is a powerful mean for maintaining the drag approximately constant as flow velocity increases.

Additionally, the shape after reconfiguration of the present bundle geometry shares some similarities with flag trees, a morphology observed when some species of trees grow in extremely windy environments. It has been recognized, \cite{telewski_2012}, that abrasion plays a role in the shaping of trees, by removing upstream branches. Using our bundle model in the case of high critical strain, typical of growing material,   shows that the upwind-facing branches would be the first to break, Fig.~\ref{fig:5}c.  This would result in a tree whose branches would be mainly oriented downwind. Such hypothesis could be validated by coupling the present results with a tree growth model.

The results shown above were obtained for a slenderness exponent $\beta$ larger than $1$, which is a necessary condition for pruning.
For beam-like plants, without tapering,  such as cereal crops and many annual plants, there can only be a single breaking event at the base since $\beta$ is then about $0$. 
This scenario is also true for young trees, which do not have enough branching levels for branch breakage to occur \cite{lopez_2011}. In this case, reconfiguration relies only on bending, and the corresponding drag curves are reduced to that computed at the highest critical strain, $\epsc=0.1$, in Figs.~\ref{fig:4} and \ref{fig:5}. Note that,  because $\beta$ does not influence how drag scales with velocity in these highly bent regimes these curves also represent the case $\beta<1$.

When $\beta$ is less than $1$, pruning is not a possible reconfiguration mechanism. It  suggests that low critical strain is not favorable in that case. 
This is consistent with the observations : crops are indeed annual plants whose critical strain is high,  and young trees have different mechanical properties than old trees, with more flexible branches and smaller Young's modulus \cite{speck_2011}. For instance, in the young walnut tree analyzed in \cite{rodriguez_2008}, $\beta$ was lower than $1$, $\beta\approx0.82$, and the Young's modulus was lower than that of older trees, suggesting a possible higher value of the critical strain.
This beneficial property could result from an evolutionary process, as proposed for perennial kelp in \cite{carrington_2013}.
The particular case of stony corals is also noteworthy, as they have a  very low critical strain, $\epsc\approx 10^{-4}$. Their geometry, where the sections of branches are  similar to that of the trunk,  suggests that breakage will occur at their base \cite{tunnicliffe_1981,*highsmith_1982,*madin_2005}. However, these particular organisms are capable of reattachment after breakage, thus ensuring their survival through breakage and dispersal. 
This is in fact a common mechanism for asexual reproduction in some terrestrial plants like \textit{Salix} and stony corals \cite{beismann_2000,tunnicliffe_1981}.
This ability to reattach can be seen as the ultimate survival strategy when neither bending nor pruning work.
The results obtained here are therefore qualitatively consistent with the observations made in nature (Table~\ref{table}). 

\begin{table}[h]
\caption{Typical survival strategy to resist large flow-induced loads and corresponding value of the critical strain $\epsc$ for different natural structures.}
\begin{ruledtabular}
\begin{tabular}{lcc}
Structure		& Strategy  		& $\epsc$\\ \hline\\[-7pt]
Corals \cite{tunnicliffe_1981}	& base  breakage and reattachment			& $10^{-4}$\\
Trees \cite{beismann_2000,speck_2011,gibson_2012}	&&\\
~~~branches	& bending/pruning and regrowth	& $10^{-2}$ \\
~~~twigs	& bending   		& $10^{-1}$	\\
Crops \cite{baker_1995}			& bending 	& $10^{-1}$	\\
\end{tabular}
\end{ruledtabular}
\label{table}
\end{table}%


In summary, different reconfiguration strategies in plants were investigated in this Letter using model geometries. We have shown how bending, pruning, or combination of the two ensures drag reduction and survival under important fluid flows. Starting from the observation that critical strains measured in nature vary by several orders of magnitude, we have focused on the effect of this mechanical parameter on reconfiguration.
It was shown that the evolution of the flow-induced drag, as the flow velocity increases, is similar for any value of the critical strain, and yields similar effects on plant survival.
In particular, when a bundle of beams with different orientations was considered, with the aim of modeling ramified plants, the drag was found to be approximatively constant as the flow velocity increases. Such a remarkable property could balance efficiently the high biological cost of pruning. These results thus suggest that the choice of either reconfiguration mechanism is not driven by survival issues, but probably related to other plant functionalities.

Hence slender organisms subjected to flow may survive to extreme loading, regardless of the critical strain of the material they are made of, by one mechanism or another, or a combination of them. This allows a large variety of states of living materials and geometries to exist in such environments, as is commonly observed \cite{koehl_1984,speck_2011}. More generally, these drag reduction strategies through shape changes may be seen as one of the posture control of plants, in reaction to an abiotic stress, see for instance in \cite{moulia_2006,bastien2013}.

%


\begin{thebibliography}{34}
\bibitem [{not()}]{note_corals}%
  \BibitemOpen
  \href@noop {} {}\bibinfo {note} {The critical strain in corals is estimated
  from values found for mineral materials like gypsum}\BibitemShut {NoStop}%
\end{thebibliography}
\end{document}